\documentclass{ws-procs975x65}

\begin{document}

\title{Freeman Dyson and Gravitational Spin Precession}

\author{N. D. Hari Dass$^*$ }

\address{Chennai Mathematical Institute, Chennai, India. \\
CQIQC, Indian Institute of Science, Bangalore, India.\\
$^*$E-mail: dass@cmi.ac.in\\
}



\begin{abstract}
In 1974 Hulse and Taylor \cite{hulse1975} discovered the binary pulsar. At that time Prof. Dyson was 
visiting the Max Planck Institute for Physics at Munich, where I was also working. He initiated a number of 
discussions on this object. During them it occurred to me that this system could be used to test Geodetic Precession in 
Einstein’s theory, which, even after years of work by the Stanford gyroscope expt \cite{stanford1960}, had remained 
a challenge. I showed some preliminary calculations to Prof Dyson and he encouraged me to do a more refined job. To be
applicable to the binary pulsar, one needed to generalise the general relativistic calculations to beyond the
so called \emph{test particle assumption}.

Barker and O'Connell \cite{barker1975a} had obtained such a result from analysing the gravitational interactions of spin-1/2 Dirac fermions
in \emph{linearized} spin-2 theories of gravitation. With C.F. Cho I produced a \emph{purely classical} calculation, using Schwinger’s 
Source theory  \cite{chodass1976}. 
B\"orner, Ehlers and Rudolf  confirmed this result 
with their general relativistic calculations shortly after \cite{BER1975}. With V. Radhakrishnan, I gave a 
detailed model for the pulse width and polarization sweep as a means of observing this effect \cite{ndhradpre1975,ndhrad1975,ndhraderr1976,
mg1975}. 
All throughout Prof. Dyson was supportive with reading the manuscripts and his critical comments. In 2005, coincidentally 
the centennial of the Annus Mirabilis(1905), Hotan, Bailes and Ord observed this in the binary pulsar J1141-6545 \cite{hotan2005}.
\end{abstract}


\bodymatter

\section{Freeman Dyson and Gravitational Spin Precession: A Tribute}
I had the great privilege of overlapping with Prof. Dyson at the Max Planck Institute for Physics in Munich during 1974-75. In October 1974,
Russell Hulse and Joseph Taylor discovered the remarkable binary pulsar PSR 1913+16, which was the most compact gravitationally bound
astrophysical system that had ever been observed \cite{hulse1975}. Prof. Dyson initiated a number of informal discussions on it. During them I got the idea
to use this system to test the spin precession predictions of Einstein's General Theory. Prof. Dyson encouraged me at every stage of
the subsequent developments. This story is a testimony not only to his great kindness to a young scientist but also to the extreme
breadth of his interests. \emph{I wish him a very happy 90th birthday.}
\section{ Gravitational spin precession}
 The most spectacular test of Einstein's general relativity theory is undoubtedly that of \emph{the bending of light}.
 The other important tests include the \emph{gravitational red shift}, the \emph{perihelion advancement} etc. 
 All these aspects had been tested with great precision.
 The theory had another remarkable prediction, namely, that of \emph{spin precession}, also called the \emph{geodetic precession}.
 It says that the axis of a gyroscope in a gravitational orbit should \emph{precess} about the orbit normal.
 In Newtonian theory there is no such effect when the gyroscope is spherically symmetric.
 The Newtonian theory does, however, predict a precession if the gyroscope is not spherically symmetric, and this is 
one of the causes of the well known 26,000 yr. \emph{precession of the equinox}.

 At the time of this work, what was known was only the spin precession equation in the \emph{test particle assumption} i.e when the gyroscope
mass is negligible compared to that of the gravitating body: 
\begin{equation}
\frac{d{\mathbf S}}{dt} = {\mathbf \omega}\times{\mathbf S}\quad\quad {\mathbf \omega} = \frac{3GM}{2R^3}\,{\mathbf R}\times{\mathbf v}
\end{equation}
As will be seen shortly, applications to the binary pulsar required the challenging task of taking this analysis beyond the test
particle assumption.
\subsection{ The Stanford gyroscope experiment}
 The idea of this experiment \cite{stanford1960} was to observe the precession of the axis of a gyroscope orbiting the earth.
 The expected rate of precession is about 6600 mas(milli-arcsec) per year.
 This was an extremely hard experiment and was not succesful in its original version. It was only in 2011 that the \emph{Gravity Probe B}
experiment succeeded in a very accurate measurement of the geodetic precession \cite{gravityB2011}.
\subsection{ The Hulse-Taylor binary pulsar}
 In October 1974, Hulse and Taylor made the remarkable discovery of a binary pulsar system \cite{hulse1975}.
 The total mass of the system was 2.6 solar masses but the orbit size was only about a solar radius. 
 The orbit period was about 8 min.(27908 s).
 The perihelion(!) motion in
this system was about 4 deg per year to be contrasted with the 46 arcsec per century value of the perihelion motion of mercury.
 It was clear that this would be an ideal place to look for many general relativistic effects including gravitational spin precession.
In fact it is only in this system that Einstein's predictions for gravitational radiation were succesfully tested.
\subsection{ Pulsars}
 Pulsars are \emph{rotating neutron stars}.
 Most observed pulsars have masses of around 1.44 solar masses.
 Because of their great masses, pulsars are gyroscopes with amazing \emph{mechanical stability}.
 Pulsars are also strongly magnetized with enormous surface fields of the order of $10^{12}$ G. 
 What makes pulsars visible(in radio waves) is the emission within narrow cones near the magnetic poles (see Fig.1).
 When our \emph{line of sight} intersects these cones, the pulsar will be visible. But due to the rapid rotation, 
the cones go out of sight, to reappear after one rotation.
 This leads to the appearance of a \emph{pulsed} radio emission to the observer.
\section{ Radhakrishnan-Cooke model}
\begin{figure}[t]
\begin{center}
\psfig{file=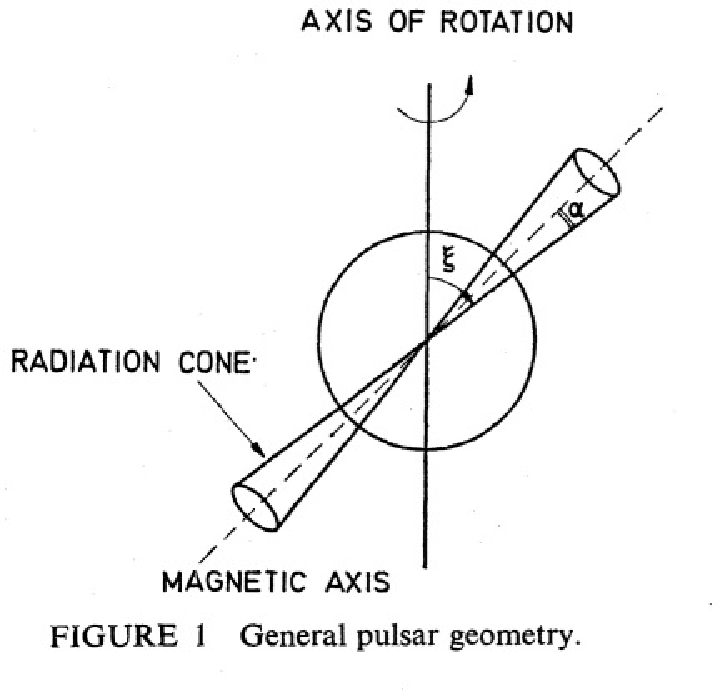,width=2in}
\end{center}
  \caption{A pulsar emission model.}
\label{aba:fig1}
\end{figure}
 The actual emission mechanism in pulsars is extremely complicated, and is not fully understood even to this day.
 Radhakrishnan and Cooke \cite{radcooke1969} proposed a very simple model which however has proved to be extremely succesful.
 According to this model, the observed radio emission is due to what they call \emph{curvature radiation}.
 Due to the extremely high surface fields, the charged particles move essentially along the magnetic field lines, and
since the latter are curved, emit the low frequency radio waves.
 The model is effectively described by only \emph{two} angular parameters, $\xi$ and $\alpha$, shown in Fig.1, where $\xi$
is the angle between the magnetic and rotation axes, and $\alpha$ is the cone angle.

 The direction of the line of sight determines both the \emph{pulse width}(it is the duration for which the line of sight is
within the cone), iand also the polarization sweep.
 The radio emission is \emph{linearly polarized}, and the \emph{instantaneous polarization} is determined by the relevant field
line.
 These two quantities i.e the pulse width and the polarization sweep are among the most important observables for a pulsar.
 Clearly, the maximum possible \emph{angular} pulse width is \emph{twice} the cone angle, and so is the maximum polarization sweep.
 These considerations play a crucial role in the observability of gravitational spin precession in the binary system.

 As mentioned before, the pulsar is a highly stable(mechanically) clock with a precision even surpassing the best \emph{atomic clocks}!
 But in the system observed by Hulse and Taylor, there was a noticeable and smooth variation in the pulse period.
 They interpreted this as due to the pulsar being in a binary, and turned the tables by using the pulse period variability to
map out the orbit very precisely and determine various orbital parameters.
 An important upshot was that the total mass of the binary system was 2.6 solar masses,
which in turn made it likely that both the objects were of \emph{comparable masses}.
 This will be seen to have crucial consequences!
\section{ Binary pulsar as a laboratory for spin precession}
 Immediately after the discovery of the binary pulsar in October 1974, Dyson, who was then on a long visit to
the Max Planck Institute at Freiman, Munich(where at that time both the Physics and Astrophysics Institutes were located), brought
this news to our attention.
 He actually initiated serious discussions stressing the importance of this object.
 At that time I, a high energy theorist immeresed in understanding the \emph{standard model}, was 
also seriously following Schwinger's source theoretic description of gravity \cite{schwingerajp1974,kimajp1974,schwingerbook1970}.
 In particular, I had been mulling about the \emph{Stanford Gyroscope Experiment} \cite{stanford1960}, inspired by the very simple and beautiful explanation
of the spin precession that Schwinger had given \cite{schwingerajp1974}.
 So quite naturally I thought that the binary pulsar, the most stable gyroscope provided by nature, should be the place to look
for this effect.
 The Stanford experiment had been facing serious technical difficulties whose main source was the lack of sufficient stability of the gyroscopes.
 So around the third week of October 1974 I made a preliminary calculation of the spin precession effect in the binary, and showed it
to Prof. Dyson. I had estimated the precession rate to be of the order of a few degrees per year as against 6600 mas for the Stanford 
experiment \cite{gravityB2011}. In fact, in the test particle limit, the spin precession rate is exactly \emph{half} the rate of
periastron advance \cite{chodass1976}.

 I had used a circular orbit for these estimates. While Prof. Dyson was very encouraging, and urged me to make a proper calculation, 
he mildly admonished me for not even using an elliptical orbit!
 While I immediately started on a proper treatment, there was one conceptual issue that was deeply \emph{troublesome}!
 The existing equations for spin precession, both in the general relativity literature, as well as in Schwinger's source description,
had been derived under the assumption that \emph{gyroscope mass is negligible}, and hence its gravitational contributions were
negligible too.
 The \emph{two body problem} in general relativity is notoriously complicated and I did not have the confidence to undertake it!
 But Schwinger's source method seemed within my grasp and I started thinking about the problem that way.

 By first week of November 1974, I went on a spectacular \emph{road journey} from Munich to Delhi via Turkey, Iran and Afghanistan,
and reached Bangalore, India, only by late December 1974.
 The binary pulsar was put on the back burner!
 In Bangalore I met Radhakrishnan, a leading expert on Pulsars, with the initial aim of suggesting the use of the large Indian radio
telescope at Ooty to make a dedicated study of the binary pulsar(it was Prof. Dyson who had stressed the desirability of this).
 I mentioned to him my preliminary calculations of the spin precession and enquired as to how one could actually observe the effects
of such a precession in pulsars.
 He was one of the creators of the Radhakrishnan-Cooke model \cite{radcooke1969}, and he explained the model to me.
 I immediately worked out the actual details of the way to observe spin precession based on this model and we wrote up our paper
on these details \cite{ndhrad1975,ndhradpre1975,ndhraderr1976}.
 An amusing coincidence here is that Radhakrishnan too was an eminent scientist without a PhD, like Prof. Dyson!
 I still had not figured a way out of the two-body difficulty, and had made the analysis for generic values of spin precession.
\subsection{Results from the Radhakrishnan-Cooke model}
In the absence of spin precession, the pulse width $\Delta T$ and the polarization sweep $\Delta \phi$, averaged over\emph{several  
orbit periods}, remain constant.
But with spin precession,  with a characterstic time scale of hundreds of years, these quantities will show a slow time
variation. A straightforward, but tedious, calculation using the Radhakrishnan-Cooke model yields:
\begin{equation}
\label{pulsewidth}
\cos\,\frac{\omega\,\Delta T(t)}{2} = \frac{\cos\alpha - \cos\xi\,\cos{\tilde \theta}_{LS}(t)}{\sin\xi\,\sin{\tilde \theta}_{LS}(t)}
\end{equation}
\begin{equation}
\label{polsweep}
\sin\,\frac{\Delta\,\phi(t)}{2} = \frac{\sin\,{\tilde \theta}_{LS}(t)}{\sin \alpha}\,\sin\,\frac{\omega\Delta T(t)}{2}
\end{equation}
\begin{equation}
\label{thetatilde}
\cos {\tilde\theta}_{LS}(t) = \cos\eta\,\cos\,\theta_{LS}+\sin\eta\,\sin\theta_{LS}\,\cos\,{\tilde\omega} t
\end{equation}
In these, $\omega$ is the angular frequency of the pulsar, ${\tilde\omega}$ is the average spin precession frequency, $\theta_{LS}$ is the angle between the line of sight and
the orbit normal, and $\eta$ is the angle between the rotation axis and the orbit normal. Rudolf Eckart carefully double checked these
results. Note that when the spin aligns itself with the orbit normal i.e $\eta=0$, there is no precession and indeed the time variations
disappear.
\section{ The two body issue.}
 By first week of March 1975 I returned to Munich and resumed thinking about the two body issue amidst finalizing the draft of
the paper with Radhakrishnan(there was no e-mail those days and letters took a week between europe and India!).
  Then I learnt of the papers by Esposito and Harrison \cite{esposito1975} on the one hand, and by Barker and O'Connell \cite{barker1975a} on the other,
addressing the spin precession issue in the binary.
 Much later on, when I made a submission of our results to the \emph{First Marcel Grossman Meeting} \cite{mg1975}, I became aware of the
paper by Damour and Ruffini, dated December 1974, also discussing the use of the binary to observe spin precession \cite{damour1974}.
 None of these had addressed the issue of the actual observation of the effect as Radhakrishnan and I had done. Damour and Ruffini had
only noted that there would be a general \emph{modulation} of the pulsar emission, while Esposito and Harrison had talked of an eventual disappearance
of the pulsar from view.

 Furthermore, the papers of Esposito and Harrison, and of Damour and Ruffini, had still used the \emph{test particle assumption} and were 
clearly not applicable to the binary. 
 Their treatments were essentially at the level where I had left in my October 1974 calculations.
 But the paper by Barker and O'Connell had a surprise!
 Though it too had not addressed the questions of actual observations, it had given the geodetic precession equations \emph{without
making the test particle assumption}!
 The new equation given by them was
\begin{equation}
\frac{d{\mathbf S}_1}{dt} = {\mathbf \omega}\times{\mathbf S}_1\quad\quad {\mathbf \omega} = \frac{G}{R^3}\,\left (2+\frac{3}{2}\frac{m_2}{m_1}\right ){\mathbf R}\times{\mathbf P}
\end{equation}
\subsection{ The Barker-O'Connell work}
 The Barker-O'Connell paper \cite{barker1975a} was based on their earlier calculation \cite{barker1975b} which was a \emph{quantum field theory} calculation in \emph{spin-2} theories of gravitation(contrary to what is stated by Hotan et al \cite{hotan2005}, this was not derived assuming general relativity to be the correct description of gravity). 
 They had studied the gravitational interactions of two spin-1/2 Dirac particles in the \emph{one-graviton exchange approximation}.
 It was expected to give, in the \emph{classical limit}, the results of general relativity.
 At that time, the precise relationship between these approaches to gravitation was still being understood.
 In particular, the classical limit, understood as $\hbar\,\rightarrow 0$, was not at all transparent.

 This was particularly acute for \emph{spin-dependent} classical gravitational effects as intrinsic spins are also of order $\hbar$.
 Even though Duff \cite{duff1973} had shown that summing tree diagrams in such a theory reproduced \emph{Scharzchild solution} in the classical limit,
no one had shown how to recover the Kerr solution likewise.
 Later on, Cho and myself showed how to understand the correctness of the Barker-O'Connell result based on \emph{low energy theorems}
for gravitation \cite{chodasseq1976}.
 The low energy theorems in fact showed that all gravitational effects of spin-2 theories had to agree with general relativity in
the \emph{large distance} limit. This was also the result obtained earlier by Hayashi \cite{hayashi1973}, and by Boulware and Deser
\cite{deser1975}.
 These results are examples of the renormalization group type arguments that David Gross \cite{grossntu} discussed in his lecture at this meeting.
\section{ Turning to Schwinger for help}
 So it became imperative to derive the precession equations beyond the test-particle approximation in a \emph{purely classical}
manner. In fact, Barker and O'Connell, in their concluding remarks \cite{barker1975b}, themselves stated explicitly:\emph{'It will be
interesting to see if these results can be derived from a purely classical treatment'}(emphasis is theirs).
 One such treatment would of course be a general relativistic calculation.
 Instead, I turned to \emph{Schwinger's Source Theory}, and recruited my friend C.F. Cho as a fellow sorcerer!
 B\"orner, Ehlers and Rudolf, at the Astrophysics institute, took the general relativity route.
 I will give a very quick guide to Schwinger's method.
 I will outline only the essence as details can be found in our papers.
\subsection{ Source Theory}
 Let us first consider the familiar case of \emph{classical electrodynamics}, but now viewed from the source perspective.
 The central object of interest is the \emph{action integral}:
\begin{equation}
W(J) = 4\pi\int\,\int\,dx\,dx^\prime\,J^\mu(x)\,D_+(x-x^\prime)\,J_\mu(x^\prime)
\end{equation}
where $J^\mu(x)$ is the total (conserved)current density, and $D_+(x-x^\prime)$ the retarded massless propagator.
 The \emph{interaction energy} is then given by
\begin{equation}
W(J) = \int\,d\tau\,E_{int}(\tau)
\end{equation}
 This suffices to determine the classical electrodynamics of a two-body system with arbitrary charges, dipole moments, quadrupole moments etc.

For the gravitational case, one starts with the action integral:
\begin{equation}
W(T) = 4\pi G\int\int \left (T_{\mu\nu}(x)T^{\mu\nu}(x^\prime)-\frac{1}{2}T(x)T(x^\prime)\right )D_+(x-x^\prime)
\end{equation}
where $T_{\mu\nu}(x)$ is the conserved energy momentum tensor.
 The interaction energy is calculated analogously.
 This suffices to determine the gravitational dynamics of a two-body system with arbitrary masses, quadrupole moments, angular momenta etc..
 The resulting spin-precession equation for the binary completely agrees with the Barker-O'Connell result, but the source calculation \cite{chodass1976} is a purely classical derivation.
 Subsequently the general relativistic calculations of B\"orner, Ehlers and Rudolf also confirmed the same \cite{BER1975}.

The source theoretic treatment by Cho and myself in fact yields the
the entire Hamiltonian of a gravitationally interacting two body system with arbitrary masses $m_1,m_2$,
spins $\vec{\bf S}_a$ and quadrupole moments ${\bf Q}_a^{ij}$. 
Details can be found in our paper. The result agrees in its physical content with the one obtained
by Barker and O'Connell, who had also obtained such a Hamiltonian from their analysis \cite{barker1975b}. Without the spin and quadrupole 
moment terms, it fully reproduces the famous Einstein, Infeld and Hoffman result \cite{einstinfeld1938}!
\section{Epilogue}
 In 2005, coincidentally the centennial of the \emph{annus mirabilis}, the spin precession effects as predicted by us
were observed by Hotan, Bailes and Ord in the binary pulsar system J1141-6545 \cite{hotan2005}. The pulsar mass in this case is
1.3 $M_{\odot}$ and the companion mass 1.0 $M_{\odot}$; the orbit period is 4.8 hrs while the periastron advance is 5.3 deg/yr. 
The expected spin precession is about 1.4 deg/yr. They observed a secular pulse broadening of 1.3 ms/yr. They estimate a spin precession rate of 0.8 deg/yr assuming $\xi=90^{\circ}$. We refer the reader to their paper for a clear account of the various observational difficulties as
well as of uncertainties in modelling. They also give an account of observed pulse broadening and estimates of spin precession rates in other
pulsars. However, the equations used by them appear to be quite different from ours, and it is not clear that they are
using the same observables as us. This needs to be understood better.
 In 2011, the final analysis of the Gravity Probe B data was made, and the result for the geodetic precession for an earth
bound gyroscope was the incredibly accurate value of \emph{6601.8$\pm$18.3 mas/yr} compared to the theoretical prediction of 6606.1
mas/yr.
 Despite its incredible accuracy, this still tests only the test particle limit of the theory.
 Prolonged observations of the effect in binary systems is the only way of testing the correct two-body aspects of the phenomenon. 
\section*{Acknowledgments}
Acknowledgments: I thank the organizers of the conference in honour of the 90th birthday of Freeman Dyson for giving me an opportunity 
to present these ideas at this meeting, and to the Institute for Advanced Study(IAS) at Nanyang Technological Institute, Singapore for
its hospitality. 
I am grateful to the Chennai Mathematical Institute and the Centre for Quantum Information and Quantum Communication(CQIQC) of the Indian Institute of Science, Bangalore for their support. I also acknowledge support from Department of Science and Technology to the project IR/S2/PU- 001/2008.
\bibliographystyle{ws-procs975x65}
\bibliography{ntu-dyson-haridass}

\end{document}